\input{aipcheck.tex}

\newcommand\sps{\space\space\space\space}
\documentclass[final]{aipproc}

\newcommand\selectedlayoutstyle{6x9}
\layoutstyle\selectedlayoutstyle

\SetInternalRegister\hbadness{8000} % pseudo latin isn't breaking very well :-)

\newcommand\doingARLO[2][]{%
  \ifx\mmref\undefined #1\else #2\fi
}

\usepackage{latexsym}
\usepackage{amsfonts}
\usepackage{amssymb}
\usepackage{amsbsy}
\usepackage{amsmath}   %needed for \begin{align}... \end{align} environment

\begin{document}

\title
      [The Field Nature of Spin]
      {The Field Nature of Spin\\ for Electromagnetic Particle}


\author{A.A. Chernitskii}{
  address={A. Friedmann Laboratory for Theoretical Physics, St.-Petersburg, Russia},
  email={AAChernitskii@engec.ru},
  altaddress={State University of Engineering and Economics, Marata str. 27, St.-Petersburg, Russia, 191002}
}

% \copyrightholder{}
\copyrightyear  {2006}

\begin{abstract}
The field nature of spin in the framework of the field electromagnetic particle concept is considered. A mathematical
character of the fine structure constant is discussed. Three topologically different field models for charged particle with spin
are investigated in the scope of the linear electrodynamics. A using of these field configurations as an initial approximation for an
appropriate particle solution of nonlinear electrodynamics is discussed.
\end{abstract}

\date{\today}

\maketitle

\section{Introduction}
The field electromagnetic particle concept in the framework of an unified nonlinear electrodynamics was discussed in my articles
(see, for example, \cite{Chernitskii1999,Chernitskii2004a,Chernitskii2006c,Chernitskii2006a}). Here I continue this theme.

\section{Electromagnetic particle with spin}

Let us consider the electromagnetic particle which is a space-localized solution
for a nonlinear electrodynamics field model. A field configuration corresponding to the solution is a
three-dimensional electromagnetic soliton.
It is not unreasonable to consider the field configuration which is more complicated
than the simplest spherically symmetrical one with point singularity. The purely Coulomb field is the known
example for such simplest configuration.
We can consider the field configuration with singly or multiply connected singular region.
This singular region can be considered to be small, so that it do not manifest explicitly in experiment.
But its implicit manifestation is the existence of the spin and the magnetic moment of the particle.

%\section{Spin}

Mass, spin, charge, and magnetic moment of the particle appear naturally in the presented approach when
the long-range interaction between the particles is considered with the help of a perturbation method
\cite{Chernitskii2006a}.
The classical equations of motion for electromagnetic particle in external electromagnetic field
are derived but not postulated. These equations are a manifestation
of the nonlinearity of the field model.  Charge and magnetic moment in this approach
characterize the particle solution at infinity. But mass and spin characterize the particle solution
in the localization region and appear as the integrated energy and angular momentum accordingly.
Thus we have the following definition for spin:
\begin{equation}
\label{63446234}
\mathtt{\mathbf{s}} = \int\pmb{\mathcal{M}}{\rm d}V
\;\;,
\end{equation}
where $\pmb{\mathcal{M}}\doteqdot\mathbf{r}\times\pmb{\mathcal{P}}$ is an angular momentum density (spin density),
$\mathbf{r}$ is a position vector,
$\pmb{\mathcal{P}}\doteqdot\,(\mathbf{D}\times\mathbf{B})/4\pi$ is a momentum density (Poynting vector).

%\section{The origin of the spin density}

\begin{figure}[h]
\label{72115434}
\put(-25,0){
\begin{picture}(30,110)
\put(74,118){$z$}
\put(90,62){$\mathbf{D}$}
\put(109,75){$\mathbf{B}$}
\put(156,62){$\pmb{\mathcal{P}}={\displaystyle\frac{1}{4\pi}}\,\mathbf{D}\times\mathbf{B}$}
\put(36,14){$\pmb{\mathcal{P}}$}
\put(39,56){$\pmb{\rho}$}
\put(-1,100){$\pmb{\mathcal{M}}_{z}=\pmb{\rho}\times\pmb{\mathcal{P}}$}
%\put(0,0){\vector(0,1){160}}
%\put(-2,80){\line(1,0){4}}
\includegraphics[height=.2\textheight]{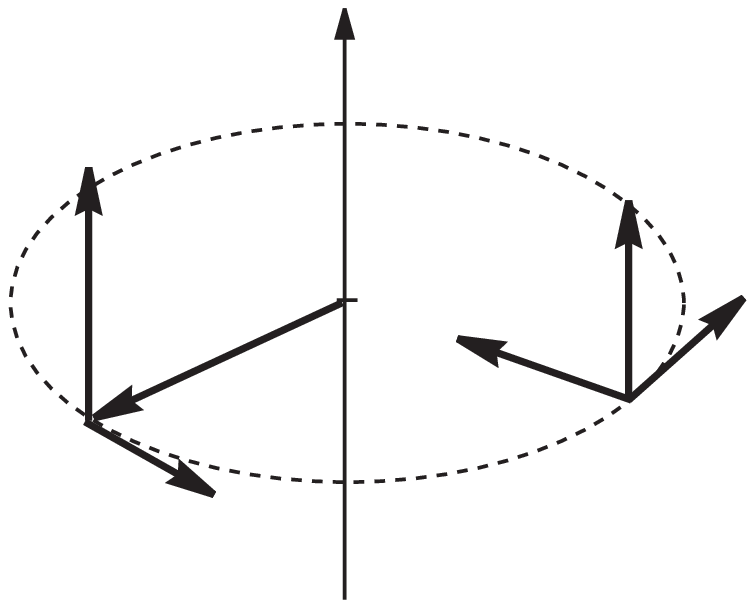}
\end{picture}
}
\caption{The origin of the spin density}
\end{figure}

The angular momentum density can appear in axisymmetric static electromagnetic field configurations with
 crossing electric and magnetic fields. In this case the crossing electric and magnetic fields give birth to the
momentum (Poynting vector) density which is tangent to a circle with center located at the axis. Because of the
axial symmetry, the full angular momentum contains only an appropriate axial component of the angular momentum density.
Thus we have the spin density directed on the axis $z$: $\pmb{\mathcal{M}}_{z}=\pmb{\rho}\times\pmb{\mathcal{P}}$,
where $\pmb{\rho}$ is a vector component of the position vector which is perpendicular to the axis $z$.
This configuration is shown on Fig. \ref{72115434}.

\section{Spin and fine structure constant}

A value of the electron spin which we have in experiment is
\begin{equation}
\label{33924859}
\mathtt{s}= \frac{\hbar}{2}= \frac{e^2}{2\,\alpha}
\quad.
\end{equation}

A particle solution of nonlinear electrodynamics, which can be appropriate to electron, has a free parameter
associated with electron charge. The existence of this free parameter is connected with a scale invariance
of the field model. Thus we have eleven free parameters for electromagnetic particle in all, viz.
ten parameters for Poincar\'{e} group and one scale parameter.

One further known continuous symmetry can be considered.
This is so called dual symmetry between electric and magnetic fields. An appropriate transformed field will have
 a magnetic charge. But this continuous transformation for the electromagnetic field be accompanied by an
 appearance of a vector potential field with a singular infinite line. Because this it is reasonable to consider that
 an appropriate free parameter does not exist for a removed particle.

We can assume that one particle solution has only these eleven free parameters.
But a separate particle included in a many-particle solution of a nonlinear field model has not these
free parameters. In the case of linear electrodynamics this separate particle has the free parameters because
of the known superposition property for the solutions. But for the case of nonlinear electrodynamics only the entire
many-particle solution has the free parameters, and an included particle has not the free parameters.

Thus we can assume that the specified value of the electron charge is connected with the nonlinearity of the model
which is the cause of the interaction between particles in the world solution.

According to formula (\ref{33924859}) we have that the dimensionless constant $2\alpha$ is the aspect ratio between the
square of the electron charge $e^2$ and the value of electron spin $\mathtt{s}$. We can consider that the electron charge is the given
constant. But the value of electron spin is calculated in the presented approach by the formula (\ref{63446234}). Thus
we can consider the fine structure constant as a mathematical one calculated by the formula
%(including Born-Infeld one)  has this symmetry property\footnote{\cite{FushchychNikitin1994}}
%\cite{FushchychNikitin1994}.
\begin{equation}
\label{72977722qqqqq}
\alpha= \frac{e^2}{2\,\mathtt{s}}
\quad,
\end{equation}
where $\mathtt{s}$ is calculated by the formula (\ref{63446234}) for the appropriate particle solution.

A value of the constant $\alpha$ calculated by the formula (\ref{72977722qqqqq}) for a particle solution in the
framework of a field model can be considered as a test for an appropriateness of these field model and particle solution.

\section{Three topologically different models for charged particle with spin}
Let us consider three topologically different models of the electron in the scope of the linear electrodynamics. These
solutions can be considered as an initial approximation to the appropriate veritable solution for a nonlinear
electrodynamics model.

%\subsection{Two-dyon configuration}
{\bf Two-dyon configuration} with equal electric and opposite magnetic charges was considered in my article \cite{Chernitskii1999}.
The spin calculated by formula (\ref{63446234}) with the appropriate solution of the linear electrodynamics does not depend
on a distance between the dyons. It is defined by the values of the electric and the magnetic charges.
If an absolute value of the ratio between the electric and the magnetic
charges (which is an additional free parameter) equals to the fine structure constant and the full charge equals to the
electron charge, then the value of spin equals to electron spin (\ref{33924859}). For the details see my article \cite{Chernitskii1999}.

%\subsection{Configuration with singular disk}
{\bf Configuration with singular disk} is obtained from the Coulomb field by a space shift with a complex number parameter
(see, for example, \cite{Burinskii1974e}).
We have the field in the complex representation for the electromagnetic field by means of the following formal complex shift:
%\begin{equation}
%\label{234729777}
${\bf{D}} + i{\bf{H}}\, = {{e{\bf{r}}}}/{{r^3 }}$,
where $x_1  \to x_1$, $x_2  \to x_2$, $x_3  \to x_3  + i\rho$.
The spin calculated by formula (\ref{63446234}) with this field does not depend on the shift parameter $\rho$ (radius of the disk).
But the numerical value of the spin for this case is not equal to the electron spin: $\mathtt{s} \approx 0.027 \,{e^2}/({2\,\alpha})$.
Note that here an additional parameter might be used to obtain the value of spin for the electron.
The magnetic part of the solution must be multiplied on this parameter.
But in this case the solution is not obtained by the complex shift from the Coulomb field.

%\subsection{Toroidal configuration}

A consideration of the static linear electrodynamics equations in toroidal coordinates $(\xi,\eta)$ gives the appropriate {\bf solution with toroidal
symmetry}. This solution can include an electric and a magnetic parts. They can be represented with the help of toroidal harmonics
which are the spheroidal harmonics with half-integer index: $P_{n-\frac{1}{2}}^l(\cosh\xi)$, where $n$ and $l$ are integer \cite{MorseFeshbachII1953}.
To obtain the right behaviour of the electromagnetic field at infinity for a charged particle with
magnetic moment we must take the toroidal harmonics $P_{-\frac{1}{2}}^0(\cosh\xi)$, $P_{-\frac{3}{2}}^0(\cosh\xi)$ for
the electric field and $P_{-\frac{1}{2}}^1(\cosh\xi)$, $P_{-\frac{3}{2}}^1(\cosh\xi)$ for
the magnetic one. Because we intend to consider this solution as an initial approximation to a solution of a nonlinear
electrodynamics model, it is reasonable to take the condition of vanishing of two electromagnetic invariants near
the singular ring. This condition will be satisfied when the ratio between the electric and magnetic vector magnitudes tends to unit
near the ring.

The spin calculated by formula (\ref{63446234}) with this toroidal field configuration does not depend on a radius of the singular ring.
But the numerical value of the spin for this case is not equal to the electron spin: $\mathtt{s} \approx 0.0006 \,{e^2}/({2\,\alpha})$.
An additional parameter might be used to obtain the value of spin for the electron.
The magnetic part of the solution must be multiplied on this parameter.
But in this case the solution does not satisfy the condition of vanishing of two electromagnetic invariants near
the ring.

\section{Conclusions and discussions}
Thus to obtain the right value of fine structure constant for the considered field configurations,
we must introduce some additional free parameter.
But, as mentioned above, for the solution of a nonlinear electrodynamics model we must have only one
free parameter that is electron charge. We must obtain the right value of the fine structure constant
in the case of an appropriate nonlinear electrodynamics model and an appropriate solution.
The examined field configurations can be considered as possible initial approximations to the right electron particle
solution.

It should be noted also that the considered here field configurations may describe
the charged particles with spin but not neutral or massless one.
However, another particle solutions may describe neutral and massless particles with spin according
to the general approach based on the formula (\ref{63446234}). These solutions may be more complicated than
considered here.

% choose bibtex style depending on layout style and options used in
% sample:

%\doingARLO[\bibliographystyle{aipproc}]
%          {\ifthenelse{\equal{\AIPcitestyleselect}{num}}
%             {\bibliographystyle{arlonum}}
%             {\bibliographystyle{arlobib}}
%          }
%\bibliography{sample}

%\renewcommand{\refname}{}\refname

%\bibliographystyle{aipproc}
%\bibliography{CHERNITSKII}

\end{document}